\documentclass[a4paper]{article}
\usepackage[utf8]{inputenc}
\usepackage{t1enc}
\usepackage[magyar,english]{babel}
\usepackage{setspace}
\usepackage{indentfirst}
\usepackage{amsmath, amsthm, amssymb}
\usepackage[pdftex]{graphicx}
\usepackage{epstopdf}
\usepackage{subfigure}
\usepackage{appendix}
\usepackage[bottom]{footmisc}
\usepackage{float}
\usepackage{cite}
\usepackage{anysize}
\usepackage{underlin}
\usepackage{fancyhdr}
\usepackage{epigraph}
\marginsize{2cm}{2cm}{2cm}{3.5cm} 
\usepackage{tikz}
\graphicspath{{./pics/}}
\usepackage{wrapfig}

\let\originalleft\left
\let\originalright\right
\renewcommand{\left}{\mathopen{}\mathclose\bgroup\originalleft}
\renewcommand{\right}{\aftergroup\egroup\originalright}

\DeclareSymbolFont{matha}{OML}{txmi}{m}{it}
\DeclareMathSymbol{\varv}{\mathord}{matha}{118}

\ifdefined\myframe
\renewenvironment{myframe}[2]{\section{#1}\begin{frame}{#2}\vspace{-10pt}}{\end{frame}} 
\else
\newenvironment{myframe}[2]{\section{#1}\begin{frame}{#2}\vspace{-10pt}}{\end{frame}}
\fi

\ifdefined\U
\renewcommand{\U}[1]{\underline{#1}}
\else
\newcommand{\U}[1]{\underline{#1}}
\fi

\ifdefined\UU
\renewcommand{\UU}[1]{\underline{\underline{#1}}}
\else
\newcommand{\UU}[1]{\underline{\underline{#1}}}
\fi

\ifdefined\mybullet
\renewcommand{\mybullet}{\vspace{2mm}\\$\bullet$ }
\else
\newcommand{\mybullet}{\vspace{2mm}\\$\bullet$ }
\fi

\ifdefined\mybulletEQ
\renewcommand{\mybulletEQ}[1]{$\bullet$ {\bf #1}\vspace{1mm}\\}
\else
\newcommand{\mybulletEQ}[1]{$\bullet$ {\bf #1}\vspace{1mm}\\}
\fi

\ifdefined\fract
\renewcommand{\fract}[2]{{\textstyle \frac{#1}{#2}}}
\else
\newcommand{\fract}[2]{{\textstyle \frac{#1}{#2}}}
\fi

\ifdefined\rect
\renewcommand{\fract}[1]{{\textstyle \frac{1}{#1}}}
\else
\newcommand{\rect}[1]{{\textstyle \frac{1}{#1}}}
\fi

\ifdefined\fracd
\renewcommand{\fracd}[2]{\frac{\displaystyle{#1}}{\displaystyle{#2}}}
\else
\newcommand{\fracd}[2]{\frac{\displaystyle{#1}}{\displaystyle{#2}}}
\fi

\ifdefined\recd
\renewcommand{\recd}[1]{\frac{\displaystyle 1}{\displaystyle{#1}}}
\else
\newcommand{\recd}[1]{\frac{\displaystyle 1}{\displaystyle{#1}}}
\fi

\ifdefined\pdd
\renewcommand{\pdd}[2]{\frac{\displaystyle{\partial{#1}}}{\displaystyle{\partial{#2}}}}
\else
\newcommand{\pdd}[2]{\frac{\displaystyle{\partial{#1}}}{\displaystyle{\partial{#2}}}}
\fi

\ifdefined\biggg
\renewcommand{\biggg}[1]{\scalebox{1.2}{\Bigg{#1}}}
\else
\newcommand{\biggg}[1]{\scalebox{1.2}{\Bigg{#1}}}
\fi

\ifdefined\Biggg
\renewcommand{\Biggg}[1]{\scalebox{1.4}{\Bigg{#1}}}
\else
\newcommand{\Biggg}[1]{\scalebox{1.4}{\Bigg{#1}}}
\fi

\ifdefined\Re
\renewcommand{\Re}{\operatorname{Re}}
\else
\newcommand{\Re}{\operatorname{Re}}
\fi

\ifdefined\Im
\renewcommand{\Im}{\operatorname{Im}}
\else
\newcommand{\Im}{\operatorname{Im}}
\fi

\ifdefined\Arch
\renewcommand{\Arch}{\operatorname{Ar\,ch}}
\else
\newcommand{\Arch}{\operatorname{Ar\,ch}}
\fi

\ifdefined\Arsh
\renewcommand{\Arsh}{\operatorname{Ar\,sh}}
\else
\newcommand{\Arsh}{\operatorname{Ar\,sh}}
\fi

\ifdefined\Arth
\renewcommand{\Arth}{\operatorname{Arth}}
\else
\newcommand{\Arth}{\operatorname{Arth}}
\fi

\ifdefined\ch
\renewcommand{\ch}{\operatorname{ch}}
\else
\newcommand{\ch}{\operatorname{ch}}
\fi

\ifdefined\sh
\renewcommand{\sh}{\operatorname{sh}}
\else
\newcommand{\sh}{\operatorname{sh}}
\fi

\ifdefined\th
\renewcommand{\th}{\operatorname{th}}
\else
\newcommand{\th}{\operatorname{th}}
\fi

\ifdefined\Ln
\renewcommand{\Ln}{\operatorname{Ln}}
\else
\newcommand{\Ln}{\operatorname{Ln}}
\fi

\ifdefined\tg
\renewcommand{\tg}{\operatorname{tg}}
\else
\newcommand{\tg}{\operatorname{tg}}
\fi

\ifdefined\ctg
\renewcommand{\ctg}{\operatorname{ctg}}
\else
\newcommand{\ctg}{\operatorname{ctg}}
\fi

\ifdefined\intl
\renewcommand{\intl}{\int\limits}
\else
\newcommand{\intl}{\int\limits}
\fi

\ifdefined\ointl
\renewcommand{\ointl}{\oint\limits}
\else
\newcommand{\ointl}{\oint\limits}
\fi

\ifdefined\integrated
\renewcommand{\integrated}[3]{\left\{{#1}\right\}\left.\vphantom{#1}\right|_{#2}^{#3}}
\else
\newcommand{\integrated}[3]{\left\{{#1}\right\}\left.\vphantom{#1}\right|_{#2}^{#3}}
\fi

\ifdefined\pd
\renewcommand{\pd}[2]{\frac{\partial{#1}}{\partial{#2}}}
\else
\newcommand{\pd}[2]{\frac{\partial{#1}}{\partial{#2}}}
\fi

\ifdefined\rec
\renewcommand{\rec}[1]{\frac{1}{#1}}
\else
\newcommand{\rec}[1]{\frac{1}{#1}}
\fi

\ifdefined\gvec
\renewcommand{\gvec}[1]{\mbox{\boldmath${#1}$}}
\else
\newcommand{\gvec}[1]{\mbox{\boldmath${#1}$}}
\fi

\ifdefined\cvec
\renewcommand{\cvec}[1]{\mbox{\boldmath${#1}$}}
\else
\newcommand{\cvec}[1]{\mbox{\boldmath${#1}$}}
\fi

\ifdefined\td
\renewcommand{\td}[2]{\frac{d{#1}}{d{#2}}}
\else
\newcommand{\td}[2]{\frac{d{#1}}{d{#2}}}
\fi

\ifdefined\md
\renewcommand{\md}[2]{\frac{\mathrm{d}{#1}}{\mathrm{d}{#2}}}
\else
\newcommand{\md}[2]{\frac{\mathrm{d}{#1}}{\mathrm{d}{#2}}}
\fi

\ifdefined\z
\renewcommand{\z}[1]{\left({#1}\right)}
\else
\newcommand{\z}[1]{\left({#1}\right)}
\fi

\ifdefined\ae
\renewcommand{\ae}[1]{\left|{#1}\right|}
\else
\newcommand{\ae}[1]{\left|{#1}\right|}
\fi

\ifdefined\sz
\renewcommand{\sz}[1]{\left[{#1}\right]}
\else
\newcommand{\sz}[1]{\left[{#1}\right]}
\fi

\ifdefined\kz
\renewcommand{\kz}[1]{\left\{{#1}\right\}}
\else
\newcommand{\kz}[1]{\left\{{#1}\right\}}
\fi

\ifdefined\B
\renewcommand{\B}[1]{\mathbb{#1}}
\else
\newcommand{\B}[1]{\mathbb{#1}}
\fi

\ifdefined\m
\renewcommand{\m}[1]{\mathrm{#1}}
\else
\newcommand{\m}[1]{\mathrm{#1}}
\fi

\ifdefined\tn
\renewcommand{\tn}[1]{\textnormal{#1}}
\else
\newcommand{\tn}[1]{\textnormal{#1}}
\fi

\ifdefined\o
\renewcommand{\o}[1]{\operatorname{#1}}
\else
\newcommand{\o}[1]{\operatorname{#1}}
\fi

\ifdefined\c
\renewcommand{\c}[1]{\mathcal{#1}}
\else
\newcommand{\c}[1]{\mathcal{#1}}
\fi

\ifdefined\v
\renewcommand{\v}[1]{\mathbf{#1}}
\else
\newcommand{\v}[1]{\mathbf{#1}}
\fi

\ifdefined\Eq
\renewcommand{\Eq}[1]{Eq.~(\ref{#1})}
\else
\newcommand{\Eq}[1]{Eq.~(\ref{#1})}
\fi

\ifdefined\Eqs
\renewcommand{\Eqs}[2]{Eqs.~(\ref{#1}) and (\ref{#2})}
\else
\newcommand{\Eqs}[2]{Eqs.~(\ref{#1}) and (\ref{#2})}
\fi

\ifdefined\a
\renewcommand{\a}[1]{\aref({#1})}
\else
\newcommand{\a}[1]{\aref({#1})}
\fi

\ifdefined\A
\renewcommand{\A}[1]{\Aref({#1})}
\else
\newcommand{\A}[1]{\Aref({#1})}
\fi

\ifdefined\r

\renewcommand{\r}[1]{(\ref{#1})}
\else
\newcommand{\r}[1]{(\ref{#1})}
\fi

\ifdefined\comm
\renewcommand{\comm}[2]{\left[{#1},{#2}\right]}
\else
\newcommand{\comm}[2]{\left[{#1},{#2}\right]}
\fi

\ifdefined\follows
\renewcommand{\follows}{\quad\Rightarrow\quad}
\else
\newcommand{\follows}{\quad\Rightarrow\quad}
\fi

\ifdefined\Follows
\renewcommand{\Follows}{\qquad\Rightarrow\qquad}
\else
\newcommand{\Follows}{\qquad\Rightarrow\qquad}
\fi

\ifdefined\followse
\renewcommand{\followse}{\quad\Rightarrow}
\else
\newcommand{\followse}{\quad\Rightarrow}
\fi

\ifdefined\bfollows
\renewcommand{\bfollows}{\Rightarrow\quad}
\else
\newcommand{\bfollows}{\Rightarrow\quad}
\fi

\ifdefined\equivalent
\renewcommand{\equivalent}{\quad\Leftrightarrow\quad}
\else
\newcommand{\equivalent}{\quad\Leftrightarrow\quad}
\fi

\ifdefined\obs
\renewcommand{\obs}[1]{\left\langle{#1}\right\rangle}
\else
\newcommand{\obs}[1]{\left\langle{#1}\right\rangle}
\fi

\ifdefined\ket
\renewcommand{\ket}[1]{\left|{#1}\right\rangle}
\else
\newcommand{\ket}[1]{\left|{#1}\right\rangle}
\fi

\ifdefined\bra
\renewcommand{\bra}[1]{\left\langle{#1}\right|}
\else
\newcommand{\bra}[1]{\left\langle{#1}\right|}
\fi

\ifdefined\braket
\renewcommand{\braket}[2]{\left<#1\vphantom{#2}\right|\left.#2\vphantom{#1}\right>}
\else
\newcommand{\braket}[2]{\left<#1\vphantom{#2}\right|\left.#2\vphantom{#1}\right>}
\fi

\ifdefined\ketbra
\renewcommand{\ketbra}[2]{\left|#1\vphantom{#2}\right>\left<#2\vphantom{#1}\right|}
\else
\newcommand{\ketbra}[2]{\left|#1\vphantom{#2}\right>\left<#2\vphantom{#1}\right|}
\fi

\ifdefined\scalprod
\renewcommand{\scalprod}[2]{\left(#1\vphantom{#2}\right|\left.#2\vphantom{#1}\right)}
\else
\newcommand{\scalprod}[2]{\left(#1\vphantom{#2}\right|\left.#2\vphantom{#1}\right)}
\fi

\ifdefined\fixmatrix
\renewcommand{\fixmatrix}[2]{\left(\begin{array}{*{9}{@{}>{\centering\arraybackslash $}m{#1}<{$ }@{}}}#2\end{array}\right)}
\else
\newcommand{\fixmatrix}[2]{\left(\begin{array}{*{9}{@{}>{\centering\arraybackslash $}m{#1}<{$ }@{}}}#2\end{array}\right)}
\fi

\ifdefined\fixgausselim
\renewcommand{\fixgausselim}[4]{\left(\hspace{-1mm}\begin{array}{*{9}{@{}>{\centering\arraybackslash $}m{#1}<{$ }@{}}}#3\end{array}\vphantom{\begin{array}{*{100}c}#4\end{array}}\hspace{-1mm}\right|\hspace{-1mm}\left.\begin{array}{*{9}{@{}>{\centering\arraybackslash $}m{#2}<{$ }@{}}}#4\end{array}\vphantom{\begin{array}{*{100}c}#3\end{array}}\right)}
\else
\newcommand{\fixgausselim}[4]{\left(\hspace{-1mm}\begin{array}{*{9}{@{}>{\centering\arraybackslash $}m{#1}<{$ }@{}}}#3\end{array}\vphantom{\begin{array}{*{100}c}#4\end{array}}\hspace{-1mm}\right|\hspace{-1mm}\left.\begin{array}{*{9}{@{}>{\centering\arraybackslash $}m{#2}<{$ }@{}}}#4\end{array}\vphantom{\begin{array}{*{100}c}#3\end{array}}\right)}
\fi

\ifdefined\gausselim
\renewcommand{\gausselim}[2]{\left(\begin{matrix}#1\end{matrix}\vphantom{\begin{matrix}#2\end{matrix}}\hspace{1mm}\right|\left.\begin{matrix}#2\end{matrix}\vphantom{\begin{matrix}#1\end{matrix}}\right)}
\else
\newcommand{\gausselim}[2]{\left(\begin{matrix}#1\end{matrix}\vphantom{\begin{matrix}#2\end{matrix}}\hspace{1mm}\right|\left.\begin{matrix}#2\end{matrix}\vphantom{\begin{matrix}#1\end{matrix}}\right)}
\fi

\ifdefined\matrixel
\renewcommand{\matrixel}[3]{\left<#1\vphantom{#2#3}\right|#2\left|#3\vphantom{#1#2}\right>} 
\else
\newcommand{\matrixel}[3]{\left<#1\vphantom{#2#3}\right|#2\left|#3\vphantom{#1#2}\right>} 
\fi

\ifdefined\contravcov
\renewcommand{\contravcov}[3]{{{#1}^{#2}_{}}_{#3}}
\else
\newcommand{\contravcov}[3]{{{#1}^{#2}_{}}_{#3}}
\fi

\ifdefined\covcontrav
\renewcommand{\covcontrav}[3]{{{#1}_{#2}^{}}^{#3}}
\else
\newcommand{\covcontrav}[3]{{{#1}_{#2}^{}}^{#3}}
\fi

\ifdefined\am
\renewcommand{\am}{{\hat{a}^{\vphantom\dagger}}}
\else
\newcommand{\am}{{\hat{a}^{\vphantom\dagger}}}
\fi

\ifdefined\ap
\renewcommand{\ap}{{\hat{a}^\dagger}}
\else
\newcommand{\ap}{{\hat{a}^\dagger}}
\fi

\ifdefined\bm
\renewcommand{\bm}{{\hat{b}^{\vphantom\dagger}}}
\else
\newcommand{\bm}{{\hat{b}^{\vphantom\dagger}}}
\fi

\ifdefined\bp
\renewcommand{\bp}{{\hat{b}^\dagger}}
\else
\newcommand{\bp}{{\hat{b}^\dagger}}
\fi

\ifdefined\arctg
\renewcommand{\arctg}{\operatorname{arctg}}
\else
\newcommand{\arctg}{\operatorname{arctg}}
\fi

\usepackage{lmodern,graphics,graphicx,amsmath,amssymb,array,hhline,color,float,cite,enumitem,url}

\begin{document}
\title{Expanded empirical formula for Coulomb final state interaction in the presence of L\'evy sources}
\author{M\'at\'e Csan\'ad$^{1}$, S\'andor L\"ok\"os$^{1,2}$ and M\'arton Nagy $^{1}$\\
$^{1}$ Eötvös Loránd University, H-1111 Budapest, P\'azm\'any P\'eter s\'et\'any 1/A\\
$^{2}$ Eszterh\'azy K\'aroly University, H-3200 Gyöngyös, Mátrai út 36.}

\maketitle

\abstract{Measurements of momentum space correlations in heavy ion reactions are a unique tools to investigate the properties of the created medium. However, these analyses require the careful handling of the final state interactions such as the Coulomb repulsion of the involved particles. In small systems such as $e^{+}+e^\textmd{--}$ or p+p the well-known Gamow factor gives an acceptable description but in the case of extended sources like that are created in heavy ion collisions, a more sophisticated approach has to be developed. In this paper we expand our previous work on the investigation of the Coulomb final state interaction in the presence of a L\'evy source. Such sources were shown to be a statistically acceptable assumption to describe the quantumstatistical correlation functions in high energy heavy ion reactions.}

\section{Introduction}

In high energy reactions, the space-time characteristics of the particle emitting source could be examined via quantumstatistical correlations. The shape of the source, hence the shape of the correlation functions are traditionally assumed to be Gaussian, but recently L\'evy distribution got much interest as a statistically acceptable description of such correlation functions in 1 and 3 dimension (for details see Ref.~\cite{PhysRevC.97.064911,Kurgyis:2018zck}). The L\'evy type source function restores the Gaussian as well as the Cauchy source function in special cases, and its parameters may carry information about underlying physical processes, such as the type of the phase transition, partial coherence or in-medium mass modifications, see e.g. ~\cite{Csorgo:2003uv,Csorgo:2005it,PhysRevC.97.064911}. Thus, the investigation of the parameters of correlation functions is crucial which implies that the precise determination of the final state interactions is desired.

The most important final state interaction is the Coulomb repulsion which can be handled with the well-known Gamow correction in small systems but in case of extended source, such as created in heavy ion collisions, it overestimates the effect. The usual approach for large systems is to take the source averaged Coulomb wave function, see e.g.~Refs.~\cite{1991PhLB..270...69B,SINYUKOV1998248}. Except some special cases, the average cannot be calculated analytically. In this paper we present our extended results on the final state Coulomb interaction in the presence of a L\'evy source. For the previous results see Ref.~\cite{Csanad:2019cns}.

\section{Coulomb effect in Bose-Einstein correlations}

In the hydrodynamical picture of the high energy collisions, the basic quantity is the source function which characterizes the particle emitting source. The one- and two-particle momentum distribution can be expressed~\cite{Yano:1978gk} with this function as
\begin{align}
N_1(\textbf{p}) = \int d^3\textbf{r} S(\textbf{r},\textbf{p}) |\psi_{\textbf{p}}(\textbf{r})|^2 \: \: \: \textmd{ and } \: \: \:
N_2(\textbf{p}_1,\textbf{p}_2) = \int d^3\textbf{r}_1 d^3\textbf{r}_2\,S( \textbf{r}_1,\textbf{p}_1)S(\textbf{r}_2,\textbf{p}_2)|\psi_{\textbf{p}_1,\textbf{p}_2}^{(2)}(\textbf{r}_1,\textbf{r}_2)|^2,
\end{align}
where $\psi_{\textbf{p}_1,\textbf{p}_2}^{(2)}$ is the two-particle wave function which must be symmetric in the spatial variables for bosons. Basically, this symmetrization effect is the origin of the (Bose-Einstein) correlations. We can introduce the pair distribution function in the following form:
\begin{align}
D(\textbf{r},\textbf{p}_1,\textbf{p}_2) = \int d^3\textbf{R} S\left(\textbf{R}+\frac{\textbf{r}}{2} \right)S\left(\textbf{R}-\frac{\textbf{r}}{2} \right).
\label{eq:Ddef}
\end{align}
Let us introduce the average and relative momentum variables as $\textbf{K} = 0.5(\textbf{p}_1+\textbf{p}_2)$ and $\textbf{q} = \textbf{p}_1 - \textbf{p}_2$. With these variables, the two-particle correlation function can be written as
\begin{align}
C_2(\textbf{q},\textbf{K})\approx \frac{\int d^3\textbf{r}\,D(\textbf{r},\textbf{K})|\psi^{(2)}_{\textbf{q}}(\textbf{r})|^2}{\int d^3\textbf{r}\,D(\textbf{r},\textbf{K})}.
\label{eq:pair_distr_corr}
\end{align}
Since the $\textbf{q}$ dependence of the correlation function is usually more rapid than the $\textbf{K}$ dependence, it is convenient to measure correlations as a function of the relative momentum at a given $\textbf{K}$ and thus the parameters will depend on $\textbf{K}$.

In the core-halo picture (Ref.~\cite{Csorgo:1994in})), the source is divided into two part: the core which contains the promptly produced particles the halo is composed by the product of resonance decays. The ratio of these parts can be characterized by the correlation strength parameter:
\begin{align}
\lambda = \frac{N_\textmd{core}}{N_\textmd{core}+N_\textmd{halo}}.
\end{align}
Assuming a source defined in Eq. \eqref{eq:Ddef}, this parameter can be introduced into the definition of the correlation function given in Eq. \eqref{eq:pair_distr_corr} as
\begin{align}
C_2(\textbf{q},\textbf{K}) = 1 -\lambda + \lambda \int d^3 \textbf{r} D_\textmd{core} (\textbf{r},\textbf{K}) |\psi^{(2)}_{\textbf{q}}(\textbf{r})|^2.
\label{eq:general_corrfunc}
\end{align}
This is the well-known Bowler-Sinyukov formula ~Refs.~\cite{1991PhLB..270...69B,SINYUKOV1998248}. One can see that the value of the correlation function evaluated at zero relative momentum is $1+\lambda$. Although the free case corresponds to $\lambda=1$, the experimental observations do not support this value. The core-halo model gives a natural explanation for this behavior of the $\lambda$ intercept parameter. Moreover, this parameter could carry information about underlying processes, such as in-medium mass modification or partial coherence, see e.g.~Refs.~\cite{Vertesi:2009wf,PhysRevC.97.064911}.

Returning to the general discussion about the Coulomb-corrected correlation function defined in Eq. \eqref{eq:general_corrfunc}, it can be seen that the evaluation of the integral requires two ingredient: the two particle source function which is assumed to be a L\'evy distribution and the two-particle Coulomb-interacting wave function. Latter can be given by the two-body scattering solution of the Schr\"odinger equation with Coulomb potential, which solution is known in the center-of-mass system of the outgoing particles (abbreviated as PCMS):
\begin{align}
\psi_{\v q}^{(2)}(\v r) = \rec{\sqrt 2}\frac{\Gamma(1{+}i\eta)}{e^{\pi\eta/2}}\kz{e^{i\v k\v r}F\big({-}i\eta,1,i(kr{-}\v k\v r)\big)
+ [\v r\leftrightarrow -\v r]},\quad\textmd{where}\quad\v k=\frac{\v q}2.
\end{align}
Here $\Gamma(\cdot)$ is the Gamma function, $F(\cdot,\cdot,\cdot)$ is the confluent hypergeometric function. The variable 
$\eta=\alpha_{\rm EM} m_\pi c/q$ is so called Sommerfeld parameter where $\alpha_{\rm EM}$ is the fine-structure constant. From now on, we restrict this analysis to pion pairs, so the mass parameter $m_\pi$ in the formula is the pion mass. Since the terminology seems not to be uniform, let us introduce the Coulomb correction as the ratio of the measured correlation function $C_2(\textbf{q})$ and the free-case correlation function $C_2^{(0)}(\textbf{q})$. We will denote the Coulomb correction with  $K(\textbf{q})$ and it can be written up according to the above convention as:
\begin{equation}
K(\textbf{q}) = \frac{C_2(\textbf{q})}{C^{(0)}_2(\textbf{q})} \: \: \: \Rightarrow \: \: \: C_2(\textbf{q}) = C^{(0)}_2(\textbf{q}) \cdot K(\textbf{q}).
\end{equation}
This form can restore the simplest case, the so-called Gamow factor that assumes the source to be a point-like particle when calculating $K(\textbf{q})$:
\begin{equation}
S(\textbf{r}) = \delta^{(3)}(\textbf{r})\: \: \: \Rightarrow \: \: \: K(\textbf{q}) = K_\textmd{Gamow}(q) = |\psi_{\textbf{q}}^{(2)}(0)|^2 = \frac{2\pi\eta}{e^{2\pi\eta}{-}1} .
\end{equation}

The result of the integral in Eq. \eqref{eq:general_corrfunc} definition cannot be expressed analytically so numerical approaches should be employed. In this paper, we present two ways to handle the Coulomb correction in a specific case of the source function -- in the presence of a L\'evy source.

\section{Lookup table for the Coulomb correction for L\'evy sources}

Recent experimental results showed that a statistically acceptable assumption for the two-particle Bose-Einstein correlation function is to utilize a L\'evy source function. The details can be found elsewhere, see e.g. Refs.~\cite{Csorgo:2003uv,PhysRevC.97.064911}. The (spherically symmetric) L\'evy distribution has two parameters, the scale parameter (radius) $R$ and the L\'evy index $\alpha$. This distribution is expressed through a Fourier transformation as 
\begin{align}
\mathcal{L}(\alpha,R,\textbf{r}) := \int\frac{d^3\textbf{q}}{(2\pi)^3}e^{i\textbf{q}\textbf{r}}\exp\left({-}\frac{1}{2}|\textbf{q}^2R^2|^{\alpha/2}\right).
\end{align}
In the $\alpha=2$ case this restores the Gaussian distribution, the $\alpha=1$ corresponds to the Cauchy distribution. For other $\alpha$ values, no simple analytic expression exists.

The integral in Eq.~\eqref{eq:general_corrfunc} cannot be evaluated analytically for L\'evy sources so it has to be calculated numerically. For experimental purposes, it is suitable to load the results into a binary file as a lookup table and than it can be used in the fitting procedure. We should interpolate between the points where the numerical calculations are actually done to express the intermediate ranges. This interpolation, however, could cause numerical fluctuations in the $\chi^2$ landscape and could mislead the fit algorithm, so an iterative procedure should be applied; ours is detailed in Ref. \cite{PhysRevC.97.064911}.

\section{Parametrization of the Coulomb correction for L\'evy sources}

In this section we discuss a different approach which is based on the numerical lookup table. As we mentioned, the finite resolution of the numerical table could cause fluctuations in the $\chi^2$ landscape and that could mislead the fit algorithm. Another disadvantage of the numerical table is its size. In practice, a sufficiently precise lookup table is slow to use. These problems can be avoided if we use a parametrization based on the table. So we get the values of the Coulomb correction from the table and parametrize its $R$ and $\alpha$ dependencies. This approach was encouraged by the successful parametrization of the $\alpha=1$ case (the Cauchy case) done by the CMS collaboration (see Ref.~\cite{Sirunyan:2017ies}, Eq. (5) for details), which was generally a corrected Gamow-correction:
\begin{align}
K_\textmd{L\'evy}(q,\alpha,R) &= K_\textmd{Gamow}(\v q)\times K_\textmd{mod}(\v q),\quad\textmd{with}\nonumber\\
K_\textmd{mod}(\textbf{q}) &= \z{1+\frac{\alpha_\textmd{EM}\pi m_\pi R}{1.26\hbar c + qR}},\quad\textmd{where}\quad
\alpha_\textmd{EM} = \frac{q_e^2}{4\pi\varepsilon_0}\rec{\hbar c}\approx\rec{137}.
\label{eq:Sikler}
\end{align}
This formula has the advantage of having only 1 numerical constant parameter (the 1.26 in the denominator) but it assumes $\alpha=1$ and we are looking for a parametrization of the general L\'evy case. We considered this as our starting point for the generalization.

\begin{figure}[h!]
\centering
\includegraphics[width=0.49\textwidth]{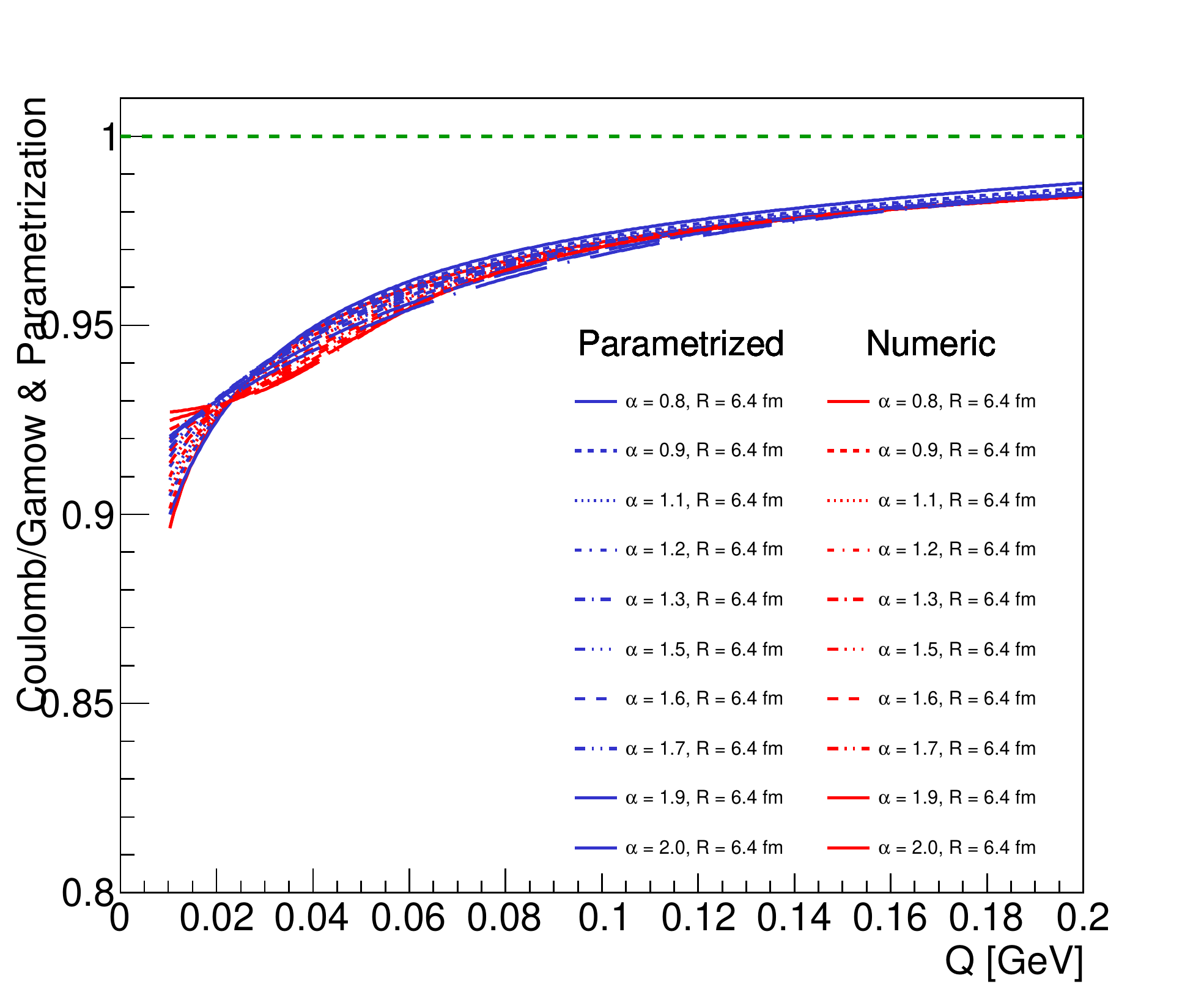}
\caption{\label{fig:basic} The fit of the Eq. \eqref{e:KLevyparam} formula (blue lines) to the values extracted from the numerical table (red lines). It can be observed that the parametrization exhibits a small deviation from the numerical calculations at the smallest $q$ values.}
\end{figure}

In order to generalize the Eq. \eqref{eq:Sikler} parametrization we chose to generalize the $K_\textmd{mod}$ correction part\footnote{The other way could be to find a completely new parametrization.} by replacing $R$ with $R/\alpha$ to introduce the $\alpha$-dependence and take higher order terms in $\frac{q R}{\alpha\hbar c}$ into consideration as it is detailed in Ref.~\cite{Csanad:2019cns}:
\begin{align}
K_\textmd{mod}(\v q) &= 1 +  \frac{A(\alpha,R)\frac{\alpha_\textmd{EM}\pi m_\pi R}{\alpha \hbar c}}
{1+B(\alpha,R)\frac{qR}{\alpha \hbar c}+C(\alpha,R)\z{\frac{q R}{\alpha \hbar c}}^2+D(\alpha,R)\z{\frac{q R}{\alpha \hbar c}}^4}.
\label{e:KLevyparam}
\end{align}
This formula simplifies to Eq.~\eqref{eq:Sikler} if $\alpha=1$ and $C=D=0$, but could follow the weak $\alpha$ dependence of the Coulomb correction (see Fig.~\ref{fig:basic}). The next step was to find suitable $A(\alpha,R)$, $B(\alpha,R)$, $C(\alpha,R)$, $D(\alpha,R)$ functions that yield acceptable approximations of the results of the numerical integration. Utilizing the previously constructed binary table, we could obtain the values of the $A, B, C, D$ functions for various values of $R$ and $\alpha$ and parametrized them with empirically found, suitable 2 dimensional functions:
\begin{align}
A(\alpha,R) &=  (a_A\alpha + a_B)^2 + (a_C R + a_D)^2 + a_E (\alpha R + 1)^2, \\
B(\alpha,R) &=  \frac{ 1 + b_A R^{b_B} - \alpha^{b_C}}{ \alpha^{2} R ( \alpha^{b_D} + b_E R^{b_F} ) }, \\
C(\alpha,R) &=  \frac{ c_A + \alpha^{c_B} +c_C R^{c_D} }{c_E} \left( \frac{\alpha}{R} \right)^{c_F},   \\
D(\alpha,R) &=  d_A + \frac{R^{d_B} + d_C \alpha^{d_F}}{R^{d_D}\alpha^{d_E}}.
\label{eq:coef_funcs}
\end{align}

The parameters in these functions are:

\begin{table}[h!]
\centering
\begin{tabular}{c c c c c c}
$a_A$ =  0.26984 & $a_B$ = -0.49123 & $a_C$ =  0.03523 & $a_D$ = -1.31628 & $a_E$ =  0.00359 & \\
$b_A$ =  2.37267 & $b_B$ =  0.58631 & $b_C$ =  2.24867 & $b_D$ = -1.43278 & $b_E$ = -0.05216 & $b_F$ = 0.72943 \\
$c_A$ = -4.30347 & $c_B$ =  0.00001 & $c_C$ =  3.30346 & $c_D$ =  0.000001 & $c_E$ =  0.000003 & $c_F$ = 1.68883 \\
$d_A$ =  0.00057 & $d_B$ = -0.80527 & $d_C$ = -0.19261 & $d_D$ =  2.77504 & $d_E$ = 2.02951 & $d_F$ = 1.07906.
\end{tabular}
\end{table}

This parametrization can follow the $R$ and $\alpha$ dependencies of the Coulomb correction in the range $0 \textmd{ GeV} / \textmd{c} < q < 0.2  \textmd{ GeV} / \textmd{c}$. As an example, for $R=6.4$ fm and different $\alpha$ values, we plotted the results of the parametrization on Fig. \ref{fig:basic}.

The functional form defined in Eq. \eqref{e:KLevyparam} cannot be used to extrapolate beyond the fitted $q$ range, deviations could appear. We can take care of these possible fluctuations with an exponential-type function which can be based on a fit to the intermediate $q$ range as $0.1-0.2$ GeV/c in the following form

\begin{align}
E(q) = 1 + A(\alpha,R)\exp\kz{ - B(\alpha,R)q},
\label{eq:exp}
\end{align}
where the $A(\alpha,R)$ and $B(\alpha,R)$ functions have a form as

\begin{align}
A(\alpha,R) &= A_a + A_b\alpha + A_c R + A_d\alpha R + A_e R^2 + A_f(\alpha R)^2,\\
B(\alpha,R) &= B_a + B_b\alpha + B_c R + B_d\alpha R + B_e R^2 + B_f(\alpha R)^2.
\end{align}
The parameters were chosen based on a fit to numerically calculated Coulomb correction at the $0.1$ GeV/c $ < q < 0.2$ Gev/c range:

\begin{table}[h!]
\centering
\begin{tabular}{c c c c c c}
$A_a$ = 0.126253 & $A_b$ = 0.05385 & $A_c$ = -0.00913 & $A_d$ = -0.01846 & $A_e$ =  0.00085 & $A_f$ = 0.00042 \\
$B_a$ = 19.31620 & $B_b$ = 5.58961 & $B_c$ =  2.26264 & $B_d$ = -1.28486 & $B_e$ = -0.08216 & $B_f$ = 0.02384.
\end{tabular}
\end{table}

This exponential damping factor should be ``smoothened together'' with the parametrization to avoid any sudden jumps. We choose to do that with a power-law-type of cut-off function:
\begin{equation}
F(q) = \frac{1}{1+\left( \frac{q}{q_0}\right)^n},
\label{eq:powerlaw}
\end{equation}
where $q_0=0.07$ GeV/c and $n = 20$. In our previous work Ref.~\cite{Csanad:2019cns}, we use an approximate Heaviside function. We investigate the difference between that case and the case we present here and we found a negligibly small difference. We still decided to improve our parametrization with the presented power-law-type of smoothing function because of its better behavior at the $q=0$ and the rapider behavior around $q_0$.

\begin{figure}[h!]
\centering
\includegraphics[width=0.48\textwidth]{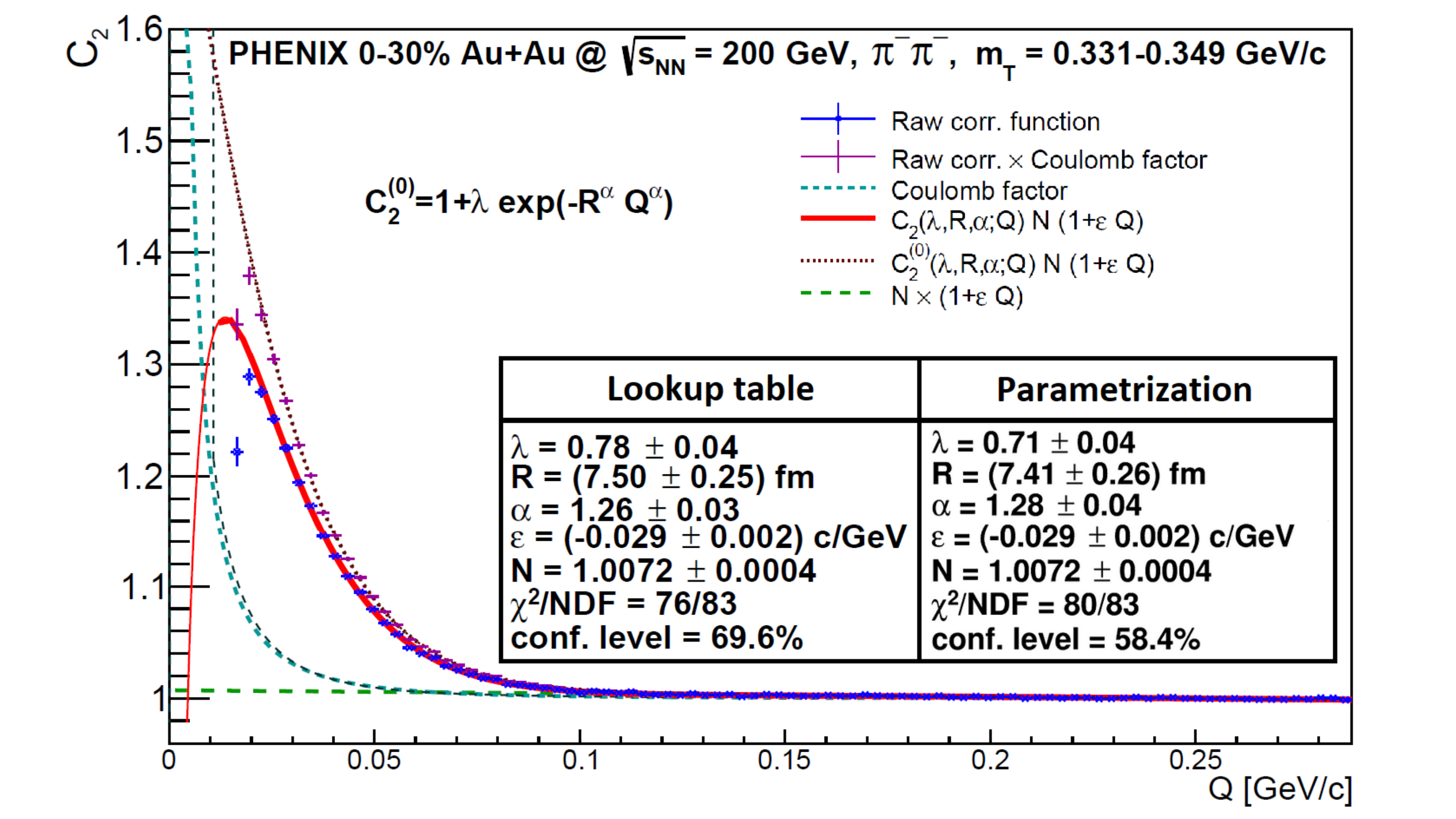}
\includegraphics[width=0.39\textwidth]{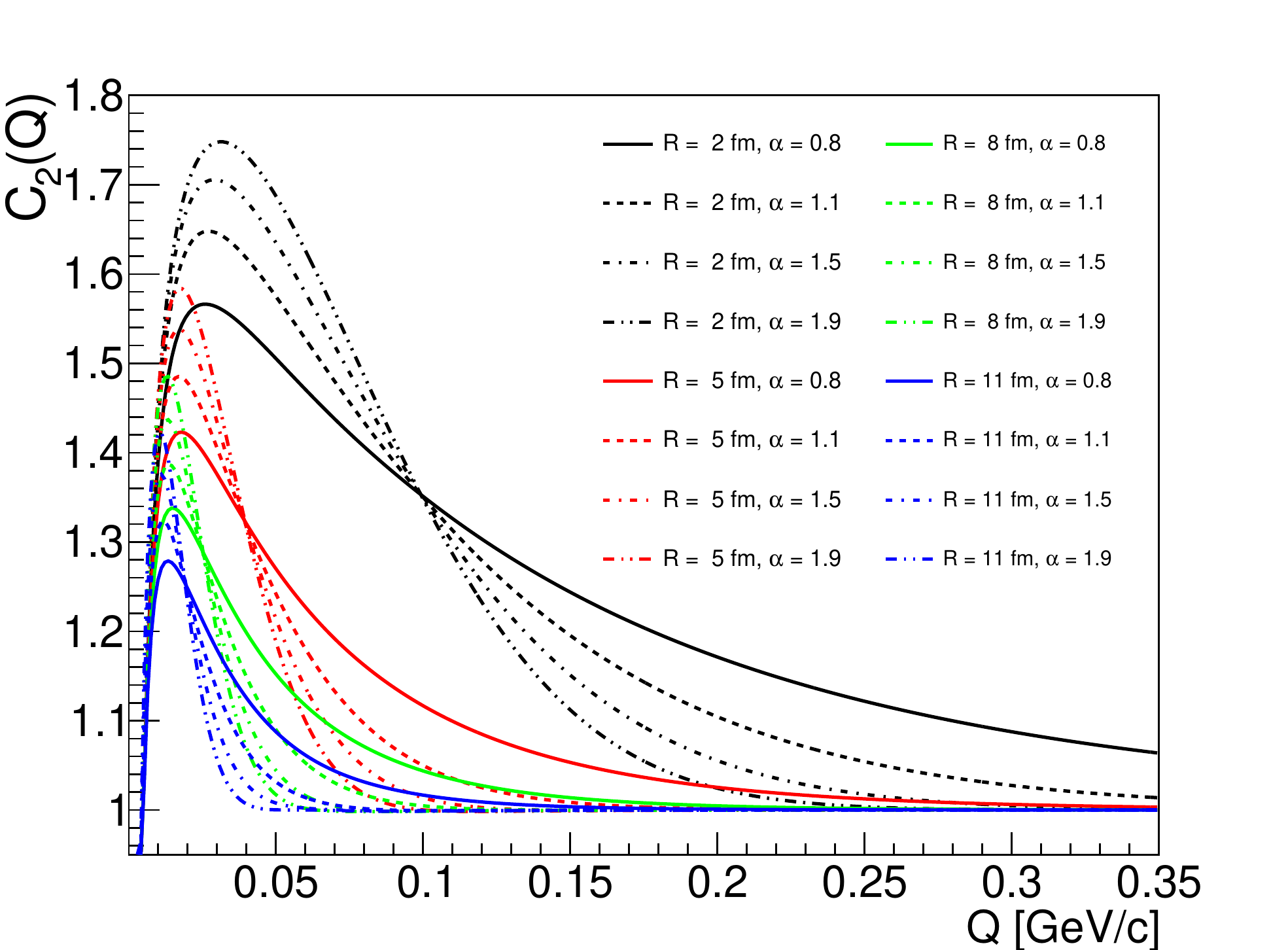}
\caption{\label{fig:repord} On the left hand side, the reproduction of previous PHENIX results~\cite{PhysRevC.97.064911} with the parametrization shown. The original PHENIX fit procedure employed the lookup numerical table, here we show our results from the parameterization. On the right hand side, we present a few example correlation functions, based on Eq.~\eqref{e:C2full}, for different $R$ and $\alpha$ values.}
\end{figure}

With the three ingredient functions defined in Eq. \eqref{e:KLevyparam}, \eqref{eq:exp} and \eqref{eq:powerlaw}, and with the corresponding parameter functions, our parametrization is valid for $0.8 < \alpha < 2$ and $2$ fm $< R < 12$ fm range. Putting everything together, we finally arrive to the form of the Coulomb correction as
\begin{equation}
K(q,\alpha,R)^{-1} =  F(q)\times K_\textmd{Gamow}^{-1}(q)\times K_\textmd{mod}^{-1}(q;\alpha,R) + (1{-}F(q))\times E(q)
\label{e:coulcorrparam}
\end{equation}
and the Coulomb corrected correlation function which could be fitted to data, according to the Bowler-Sinyukov method, can be written in the form of
\begin{equation}
C_2(q;\alpha,R) = \left[ 1 - \lambda + K(q;\alpha,R) \lambda \left( 1 + \exp \left[ | q R |^{\alpha} \right] \right) \right] \cdot (\textmd{assumed background}).
\label{e:C2full}
\end{equation}

We used this formula to reproduce previous PHENIX results from Fig. 3. of Ref.~\cite{PhysRevC.97.064911}~\footnote{The data of the shown PHENIX correlation function result can be found at \url{https://www.phenix.bnl.gov/phenix/WWW/info/data/ppg194_data.html}.}; this can be seen on Fig.~\ref{fig:repord}. The two fits are compatible with each other. An example code calculating the Coulomb correction as defined in Eq. (\ref{e:coulcorrparam}) can be found in Ref.~\cite{coulcorrparamcode}.

\begin{figure}[h!]
\centering
\includegraphics[width=0.45\textwidth]{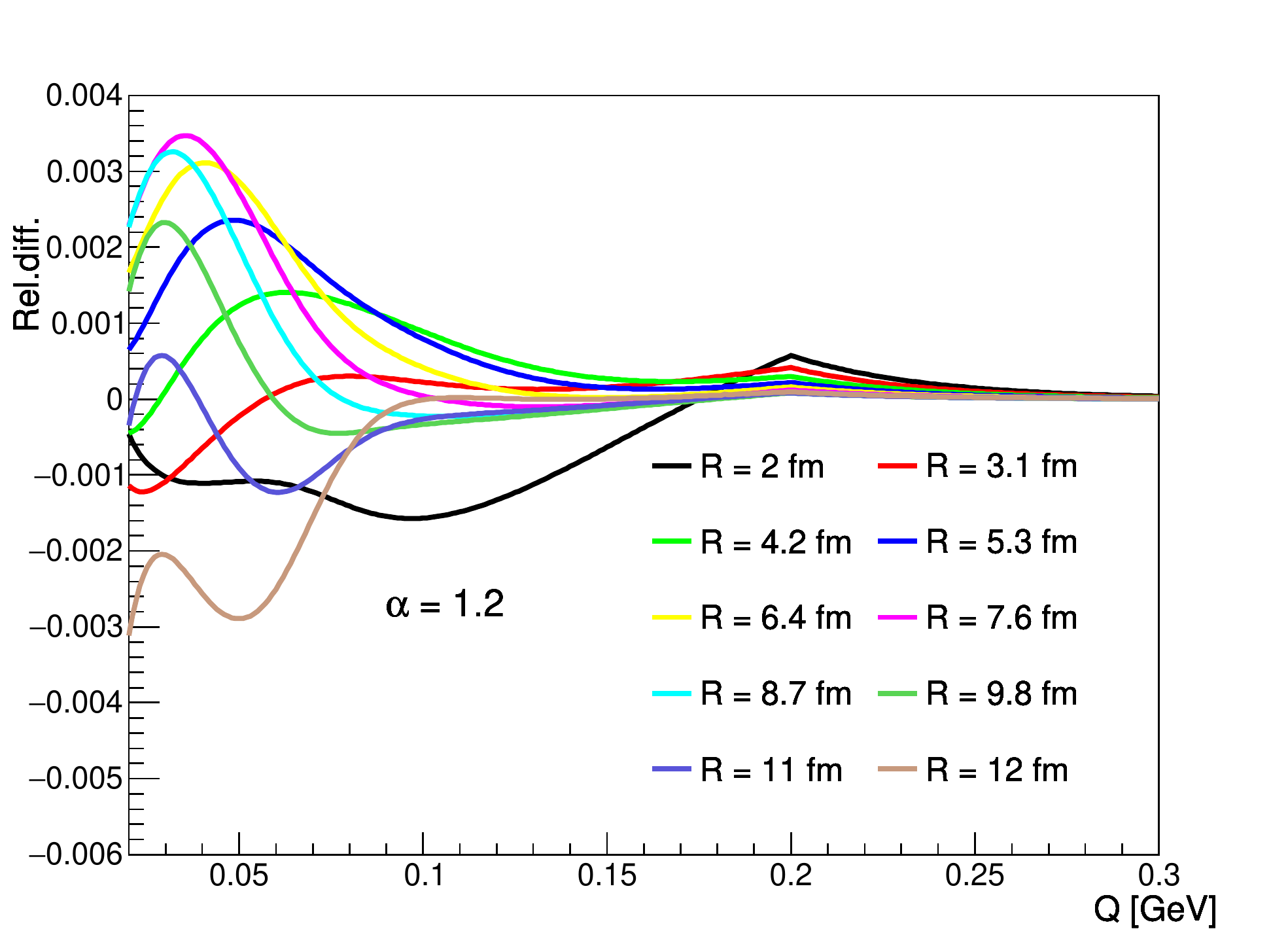}
\includegraphics[width=0.45\textwidth]{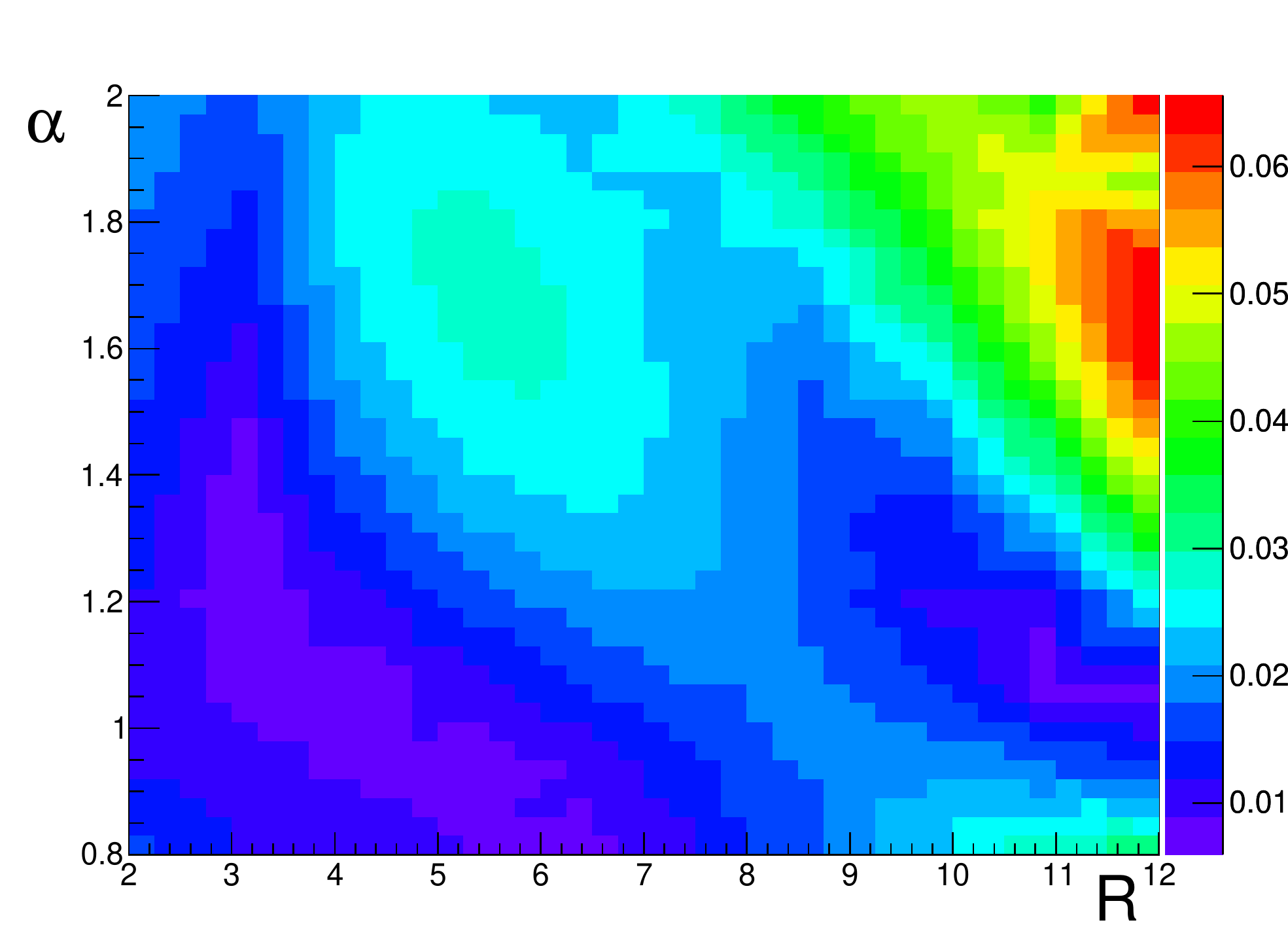}
\caption{\label{fig:reldev} On the left hand side we present the relative deviation of the parametrization form the table measured in $\%$ for  $\alpha=1.2$ with various $R$ values. On the right hand side the $q$-averaged relative difference is presented. (Averaging was done on $0.01 < q < 0.1$ GeV/$c$.)}
\end{figure}

We investigated the parametrization by looking at its relative deviation from the lookup table also in the case when $\alpha=1.2$ with different $R$ values and with a two-dimensional histogram of the relative differences in in Fig.~\ref{fig:reldev}. The maximum of these relative differences is around 0.07\%.

\section{Conclusions}

We presented our results on the effect of the Coulomb repulsion to Bose-Einstein correlations in high energy heavy ion reactions in the presence of a L\'evy source. We investigate two equivalent method which could be used in experimental practice. One of them is to fill the values of the numerical integral of the Coulomb correction into a binary table, the other one is a parametrization based on the table. The described parametrization is valid for $R=2-12$ fm and $\alpha = 0.8-2.0$ ranges, and shown to be compatible with previous results and with the numerical table. Thus, our parametrization can be used in quantumstatistical correlation measurements effectively that assume Cauchy, Gaussian or the more general L\'evy sources.


\section*{Acknowledgments}This work was supported by the NKFIH grant FK 123842. S.L. is grateful for the support of EFOP 3.6.1-16-2016-00001. M.N. and M.Cs. are supported by the Hungarian Academy of Sciences through the ``Bolyai J\'anos'' Research Scholarship program as well as the \'UNKP-19-4 New National Excellence Program of the Hungarian Ministry of Technology and Innovation. The authors would like to thank B. Kurgyis for the careful reading of our manuscript and the useful suggestions.

\end{document}